# Energy Density Bands Engineering

A. D. Alhaidari
*Saudi Center for Theoretical Physics, Jeddah, Saudi Arabia*

**Abstract**: We expand the quantum mechanical wavefunction in a complete set of square integrable orthonormal basis such that the matrix representation of the Hamiltonian operator is tridiagonal and symmetric. Consequently, the matrix wave equation becomes a symmetric three-term recursion relation for the expansion coefficients of the wavefunction whose solution is a set of orthogonal polynomials in the energy. The polynomials weight function is the energy density of the system constructed using the Green's function, which is written in terms of the Hamiltonian matrix elements. We study the distribution of zeros of these polynomials on the real energy line based exclusively on their three-term recursion relations. We show that the zeros are generally grouped into sets belonging to separated bands on the orthogonality interval. The number of these bands is equal to the periodicity (multiplicity) of the asymptotic values of the recursion coefficients and the location of their boundaries depend only on these asymptotic values. Bound states (if they exist) are located at discrete zeros found in the gaps between the density bands that are stable against variation in the order of the polynomial for very large orders. We give examples of systems with a single, double and triple energy density bands.

## 1. Introduction:

The energy density of states $\rho(E)$, which is the subject of our study here, is defined as the number of continuum states per unit energy. We consider the normalized version where its integral over all available energies is one. Complex systems have rich energy density structures where, for example, $\rho(E)$ consists of several bands with gaps in between where the density vanishes (aside from a finite number of discrete points that correspond to bound states). If the system is driven (excited) at high enough energy then it will experience spontaneous transition from states in one band to the other jumping over the forbidden energy gaps. Additionally, the shape of the individual energy bands could be highly non-trivial. The structure of $\rho(E)$ is very important for designing and manufacturing materials with specific electrical, magnetic, thermal, optical and mechanical properties [1]. The construction of such density of states is what we mean by "*energy density bands engineering*". Usually, the task comes down to searching for a highly complex structured system or equivalently elaborate potential function that models such a system. Nonetheless, in our Tridiagonal Representation Approach (TRA), which we use to obtain exact algebraic solutions of the wave equation [2], we do have a viable and simpler alternative for constructing such complex systems. In the process, we take full advantage of the connection between tridiagonal matrices and orthogonal polynomials, continued fractions, and quadrature approximation [3]. The proposed method parallels that of the "*recursion method*" [4-8] but with emphasis on the associated orthogonal polynomials and their properties (e.g., weight function, distribution of zeros, spectral property, etc.). Thus, the construction here is more algebraic than analytical making the mathematical labor less intense and gives wider applicability due to the empirical nature of the proposed method. Our study here is structural where we focus on the shape of the density of states and its band composition but we will not address the associated dynamics (e.g., transition and/or tunneling between the bands, etc.).

Now in the TRA, the wavefunction is written as the infinite bounded sum $\left|\psi^{\mu}(E,x)\right\rangle = \sum_n f_n^{\mu}(E)\left|\phi_n(x)\right\rangle$, where $\{\phi_n(x)\}$ is a complete set of square integrable basis functions, $\left\{f_n^{\mu}(E)\right\}$ are appropriate expansion coefficients in the energy and $\{\mu\}$ is a set of real physical parameters. We require that the basis elements be chosen such that the matrix representation of the wave operator is tridiagonal and symmetric. Other than this requirement and that they satisfy the boundary conditions, the basis functions contain no physical information about the system. All structural and dynamical information are contained in the expansion coefficients. Next, we write $f_n^{\mu}(E) = f_0^{\mu}(E)P_n^{\mu}(E)$, where $\left|f_0^{\mu}(E)\right|^2$ is the energy density of states of the system [2,9]. This also makes $P_0^{\mu}(E)=1$. From this point on, we assume that the basis elements are orthonormal (i.e., $\langle\phi_n|\phi_m\rangle = \delta_{nm}$). Consequently, the TRA requirement that the wave operator matrix, $\langle\phi_n|(H-E)|\phi_m\rangle$, be tridiagonal and symmetric dictates that the Hamiltonian matrix is so, too. That is, we can write the matrix representation of the Hamiltonian as follows

$$H = \begin{pmatrix} a_0 & b_0 & & & & & \\ b_0 & a_1 & b_1 & & \mathbf{0} & & \\ & b_1 & a_2 & b_2 & & & \\ & & b_2 & a_3 & b_3 & & \\ & & & \times & \times & \times & \\ & \mathbf{0} & & & \times & \times & \times \\ & & & & & \times & \times \end{pmatrix} \tag{1}$$

where $\{a_n, b_n\}$ are real constants that depend on $\{\mu\}$ and such that $b_n \neq 0$ for all *n*. Therefore, the matrix wave equation, $\langle\phi_n|(H-E)|\psi^{\mu}(E,x)\rangle = 0$, results in the following symmetric three-term recursion relation

$$E\,P_n^{\mu}(E) = a_n P_n^{\mu}(E) + b_{n-1}P_{n-1}^{\mu}(E) + b_n P_{n+1}^{\mu}(E)\,. \tag{2}$$

Hence, $P_n^{\mu}(E)$ is a polynomial in *E* of order *n*. The recursion (2) is valid for $n = 1,2,3,...$ together with the initial seeds $P_0^{\mu}(E)=1$ and $P_1^{\mu}(E) = (E-a_0)/b_0$. These are referred to as "*polynomials of the first kind*". The "*polynomials of the second kind*", $\left\{\hat{P}_n^{\mu}(E)\right\}$, satisfy the same recursion relation (2) but with $\hat{P}_0^{\mu}(E)=1$ and $\hat{P}_1^{\mu}(E) = c_0 + c_1 E$ where the linearity constants $c_0$ and/or $c_1$ differ from those of the first kind. Normalization of the wavefunction in space and energy together with the completeness of the basis, $\sum_n \phi_n(x)\phi_n(y) = \delta(x-y)$, imply that

$$\int \left|f_0^{\mu}(E)\right|^2 P_n^{\mu}(E)P_m^{\mu}(E)dE = \delta_{nm}\,. \tag{3}$$

Consequently, $\left\{P_n^{\mu}(E)\right\}$ becomes a complete set of orthonormal polynomials whose positive definite weight function is the energy density of states $\rho^{\mu}(E) = \left|f_0^{\mu}(E)\right|^2$. Taking $n=m=0$ in (3) shows that this energy density is normalized to one. The Green's function associated with this system is defined as $G(z) = (H-z)^{-1}$ where *H* is the infinite tridiagonal Hamiltonian matrix (1). The matrix elements of this Green's function is

$$G_{nm}(z)=\langle\phi_n|(H-z)^{-1}|\phi_m\rangle=\int\rho^{\mu}(E)\frac{P_n^{\mu}(E)P_m^{\mu}(z)}{E-z}dE\,, \tag{4}$$

where the integral runs over all allowed energies. Taking $n=m=0$ and since $P_0^{\mu}=1$, then Cauchy's integral theorem shows that we can define the energy density of states $\rho^{\mu}(E)$ as the discontinuity of $G_{00}(z)$ across the cut along the real line in the complex energy plane. That is, we write

$$\rho^{\mu}(E)=\frac{1}{2\mathrm{i}\pi}\lim_{\varepsilon\to 0}\left[G_{00}(E+\mathrm{i}\varepsilon)-G_{00}(E-\mathrm{i}\varepsilon)\right]=\frac{1}{\pi}\lim_{\varepsilon\to 0}\left\{\mathrm{Im}\left[G_{00}(E+\mathrm{i}\varepsilon)\right]\right\}, \tag{5}$$

Using one of its representations, we can write $G_{00}(z)$ as the following infinitely continued fraction [4,10]

$$G_{00}(z)=\cfrac{-1}{z-a_0-\cfrac{b_0^2}{z-a_1-\cfrac{b_1^2}{z-a_2-\cfrac{b_2^2}{z-a_3-...}}}} \tag{6}$$

One can also show that the density of states associated with the system of polynomials of the second kind, $\hat{\rho}^{\mu}(E)$, is related to $\rho^{\mu}(E)$ as follows

$$\hat{\rho}^{\mu}(E)=\frac{\rho^{\mu}(E)}{\left|1+b_0\left[P_1^{\mu}(E)-\hat{P}_1^{\mu}(E)\right]G_{00}(E+\mathrm{i}0)\right|^2}. \tag{7}$$

This could be obtained using the following process: First, we allow the recursion relation (2) to be true for $n=0,1,2,...$ by defining $b_{-1}\equiv 0$ while taking the zero order polynomial equal to one as usual. Second, we modify the recursion relation by adding the term $b_0[P_1^{\mu}(E)-\hat{P}_1^{\mu}(E)]\delta_{n0}$ to the right-hand side of the recursion relation. Finally, we derive the new $\hat{G}_{00}(z)$ in terms of the original $G_{00}(z)$ and subsequently obtain $\hat{\rho}^{\mu}(E)$ in terms of $\rho^{\mu}(E)$.

Now, we state the main findings of the work to be derived either analytically or empirically below and with examples. We start by assuming that the asymptotic limit ($n\to\infty$) of the recursion coefficients $\{a_n,b_n\}$ is multivalued and as follows:

$$\lim_{n\to\infty}\{a_n,b_n\}=\{A_k,B_k\}_{k=1}^{K}, \tag{8}$$

where *K* is a natural number. This means that the tail of the tridiagonal Hamiltonian matrix (1) oscillates between *K* set of fixed values. Armed with rich experience acquired from dealing with many such systems of different Hamiltonian matrices like (1) that satisfy (8) and for various values of *K*, we can state the following conclusions:

1. The energy density of states $\rho(E)$ consists of *K* separated energy bands on the real energy line.
2. The 2*K* boundaries of the energy bands depend only on the asymptotics $\{A_k,B_k\}_{k=1}^{K}$. For $K=1,2,$ and 3, these boundaries are obtained as follows:
   $K=1$: The two boundaries of the single band are located at $E=A_1\pm 2B_1$.
   $K=2$: The four boundaries of the two bands are located at:

$$E=\frac{A_1+A_2}{2}+\frac{1}{2}\sqrt{\left(A_1-A_2\right)^2+4\left(B_1\pm B_2\right)^2}\,,$$
$$E=\frac{A_1+A_2}{2}-\frac{1}{2}\sqrt{\left(A_1-A_2\right)^2+4\left(B_1\pm B_2\right)^2}\,.$$

$K=3$: The six boundaries of the three bands are obtained as real solutions of the equation: $\beta^2(z)-4\alpha(z)\gamma(z)=0$, where $\alpha(z)=B_3^2\left[B_1^2-F_1(z)F_2(z)\right]$, $\beta(z)=B_2^2F_1(z)-B_3^2F_2(z)+B_1^2F_3(z)-F_1(z)F_2(z)F_3(z)$, $\gamma(z)=B_2^2-F_2(z)F_3(z)$, and $F_k(z)=z-A_k$.

3. All zeros of the polynomials $\left\{P_n^\mu(E)\right\}$ lie within the energy bands except for possibly a maximum number of *K* zeros that are located within the $K-1$ gaps between the bands.
4. The zeros in the gaps that are stable against variation in the polynomial order *n* as $n\to\infty$ correspond to discrete bound states. Thus, the number of such bound states are less than or equal to *K*.

The last finding is evident since it means that at those special energies (zeros of the polynomials) the asymptotic wavefunction vanishes as required for a bound state. Next, we prove the first two findings analytically. Now, since all $\{a_n,b_n\}$ are real, then terminating the continued faction (6) at any finite order no matter how large will results only in real values for the $\lim\limits_{\varepsilon\to 0}\left[G_{00}(E+i\varepsilon)\right]$ leading to zero density of states. This dictates that one should evaluate $G_{00}(z)$ in the complex energy plane then make an analytic continuation to the real line from above and from below. The discontinuity will then give the energy density. Nonetheless, we follow a simpler and effective approach by making the following proper approximation of $G_{00}(z)$:

$$G_{00}(z)=\cfrac{-1}{z-a_0-\cfrac{b_0^2}{z-a_1-\cfrac{b_1^2}{z-a_2-\ldots-\cfrac{b_{N-2}^2}{z-a_{N-1}+b_{N-1}^2T(z)}}}} \tag{9}$$

where *N* is some large enough integer such that $\{a_n,b_n\}_{n\geq N}$ are as close as desired to the asymptotic values $\{A_k,B_k\}_{k=1}^K$. This means that the "*Terminator Function*" *T*(*z*) is

$$T(z)=\cfrac{-1}{z-A_1-\cfrac{B_1^2}{z-A_2-\cfrac{B_2^2}{z-A_3-\ldots.-\cfrac{B_{K-1}^2}{z-A_K-\cfrac{B_K^2}{z-A_1-\cfrac{B_1^2}{z-A_2-\cfrac{B_2^2}{z-A_3-\ldots..}}}}}}} \tag{10}$$

which could be rewritten as follows

$$T(z)=\cfrac{-1}{z-A_1-\cfrac{B_1^2}{z-A_2-\cfrac{B_2^2}{z-A_3-.....-\cfrac{B_{K-1}^2}{z-A_K+B_K^2T(z)}}}} \tag{11}$$

It is obvious that this gives a quadratic equation for $T(z)$ whose general form is $\alpha(z)T^2+\beta(z)T+\gamma(z)=0$. By the process of induction, one can show that $\beta(z)$ is a polynomial in $z$ of order $K$ whereas both $\alpha(z)$ and $\gamma(z)$ are polynomials of order $K-1$ in $z$. Therefore, the solution of this quadratic equation for $T(z)$ becomes complex if the discriminant $\beta^2(z)-4\alpha(z)\gamma(z)$ becomes negative for certain range(s) of $z$. Thus, $G_{00}(z)$ is complex only in these ranges giving the sought after non-zero density of states in these energy intervals (bands). The boundaries of the bands are located at energies that make the discriminant vanish. Now, the discriminant is a polynomial in $z$ of order $2K$, which is obtained from the special $K^{\text{th}}$ order continued fraction (11) associated with the following tridiagonal symmetric matrix

$$\begin{pmatrix} A_1 & B_1 & & & & & \\ B_1 & A_2 & B_2 & & & \mathbf{0} & \\ & B_2 & A_3 & B_3 & & & \\ & & \times & \times & \times & & \\ & & & \times & \times & \times & \\ & \mathbf{0} & & & B_{K-2} & A_{K-1} & B_{K-1} \\ & & & & & B_{K-1} & A_K \end{pmatrix}. \tag{12}$$

Thus, in principle, the discriminant polynomial will have $2K$ real roots as boundaries of the $K$ energy density bands. That is, the boundaries of the bands are located at the $2K$ real roots of the polynomial $\beta^2(z)-4\alpha(z)\gamma(z)$ and thus these boundaries are uniquely determined by the asymptotic coefficients $\{A_k,B_k\}_{k=1}^{K}$. In the following three sections, we treat the cases corresponding to $K=1$, 2 and 3 separately. Additionally, we assume that the energy is measured in some universal units such that we can take the entries in the Hamiltonian matrix (1) to be dimensionless.

## 2. The case: *K* = 1

For $K=1$, we obtain $T(z)=\dfrac{-1}{z-A_1+B_1^2T(z)}$ giving

$$T(E)=\frac{1}{2B_1^2}\left[A_1-E\pm\sqrt{(E-A_1-2B_1)(E-A_1+2B_1)}\right]. \tag{13}$$

This is a real expression except if $E\in\left[A_1-2B_1,A_1+2B_1\right]$. Therefore, substituting (13) into (9) then into (5) will result in non-zero density of states as a single band whose boundaries are at $E=A_1\pm 2B_1$. The shape of this band depends on the specific form of the recursion coefficients. For a single energy density band, the choice of sign in front of the square root in (13) is system dependent. However, as we shall see below, for multi-band densities this choice alternates from one band to the next. As an example of a single band energy density, we consider the system whose Hamiltonian matrix (1) is defined by

$$a_n = \gamma\left(\delta_{n,0} + \delta_{n,1}\right) \text{ and } b_n = \tfrac{1}{2}\left(\tfrac{n+\alpha}{n+\alpha^{-1}}\right)^{\beta}, \tag{14}$$

where $\alpha$ is positive. Thus, $A_1 = 0$ and $B_1 = \frac{1}{2}$. Figure 1a is a plot of the energy density obtained for a given set of physical parameters $\{\alpha,\beta,\gamma\}$ and for large enough integer *N*. Figure 1b is a set of snap shots from the animation that shows the distribution of zeros (small filled circles) of the polynomials that satisfy the recursion relation (2) with recursion coefficients (14) and seeds: $P_0^{\mu}(E) = 1$ and $P_1^{\mu}(E) = (E - a_0)/b_0$.

## 3. The case: *K* = 2

For *K* = 2, we obtain $T(z) = -\left[z - A_1 - \dfrac{B_1^2}{z - A_2 + B_2^2 T(z)}\right]^{-1}$ giving

$$2B_2^2 T(E) = \frac{1}{E - A_1}\left\{B_1^2 - B_2^2 - (E-A_1)(E-A_2) \pm \sqrt{\left[(E-A_1)(E-A_2) - (B_1 - B_2)^2\right]\left[(E-A_1)(E-A_2) - (B_1+B_2)^2\right]}\right\} \tag{15}$$

This is a real expression except if the two brackets under the square root have different signs. The square root vanishes at the following real four energy points

$$E = \frac{A_1 + A_2}{2} + \frac{1}{2}\sqrt{\left(A_1 - A_2\right)^2 + 4\left(B_1 \pm B_2\right)^2}\text{ , and} \tag{16a}$$

$$E = \frac{A_1 + A_2}{2} - \frac{1}{2}\sqrt{\left(A_1 - A_2\right)^2 + 4\left(B_1 \pm B_2\right)^2}\text{ .} \tag{16b}$$

Sorting these four energy points as $E_1 < E_2 < E_3 < E_4$ enables us to rewrite (15) as

$$2B_2^2 T(E) = \frac{1}{E - A_1}\left[B_1^2 - B_2^2 - (E-A_1)(E-A_2) \pm \sqrt{(E-E_1)(E-E_2)(E-E_3)(E-E_4)}\right] \tag{15'}$$

Therefore, *T*(*E*) is real except if $E_1 < E < E_2$ or $E_3 < E < E_4$. Substituting this terminator into (9) then into (5) results in non-zero density of states as a double band on the two energy intervals $[E_1, E_2]$ and $[E_3, E_4]$ with a gap in the range $[E_2, E_3]$. Note that the sign in front of the square root in (15) or (15)' for one density band is opposite to that of the other band. All zeros of the associated polynomials fall within the two density bands except for possibly a maximum of two that lie within the gap. This is shown explicitly in the second example where

$$a_{2n} = \gamma,\ a_{2n+1} = 1 - \gamma,\ b_{2n} = \frac{\beta}{2}\left[\frac{n+\alpha^{-1}}{n+\alpha}\right]^{\gamma},\text{ and } b_{2n+1} = \frac{\alpha}{2}\left[\frac{n+\beta}{n+\beta^{-1}}\right]^{1-\gamma}, \tag{17}$$

and where $\{\alpha,\beta,\gamma\}$ is a set of dimensionless positive parameters. Thus, $\{A_1, A_2\} = \{\gamma, 1-\gamma\}$ and $\{B_1, B_2\} = \frac{1}{2}\{\beta,\alpha\}$. Figure 2a is a plot of the energy density obtained by substituting (15) into (9) then into (5) for a given set of physical parameters $\{\alpha,\beta,\gamma\}$ and for large enough integer *N*. Figure 2b is a set of snap shots from the animation that shows the distribution of zeros of the polynomials of the first kind that satisfy the recursion relation (2). The zeros are the small filled circles in the figure. Note the presence of a single zero for odd polynomials inside the gap and very close to the left side of the gap. Since it is asymptotically ($n \to \infty$) stable then it represents the bound state energy for odd parity states only. That is, the asymptotic wavefunction at this energy will not have odd states (i.e., states that span the $\{\phi_{2n+1}(x)\}$ basis elements). However, it does not represent a bound state for the whole system since it is not present in all polynomials (odd and even). Had the energy gap been wide enough, then it would have been possible for two zeros to fall within the gap and if any one of them is shared by odd and even polynomials and is asymptotically stable, then that would have been a bound state energy for the whole system.

## 4. The case: *K* = 3

Following the same procedure as in the above two cases but with somewhat lengthier algebra, we obtain the terminator as solution of the quadratic equation $\alpha(z)T(z)^2+\beta(z)T(z)+\gamma(z)=0$, where $\alpha(z)=B_3^2\left[B_1^2-F_1(z)F_2(z)\right]$, $\beta(z)=B_2^2F_1(z)-B_3^2F_2(z)+B_1^2F_3(z)-F_1(z)F_2(z)F_3(z)$, $\gamma(z)=B_2^2-F_2(z)F_3(z)$ with $F_k(z)=z-A_k$. Therefore, the terminator function reads as follows

$$T(z)=\frac{1}{2\alpha(z)}\left[-\beta(z)\pm\sqrt{\beta(z)^2-4\alpha(z)\gamma(z)}\right]. \tag{18}$$

Substituting this terminator into (9) then into (5) results in non-zero three-band density of states with two energy gaps. The sign in front of the square root in (18) alternates among the bands from left to right. The boundaries of the three density bands are the six real roots of the polynomial $\beta^2(z)-4\alpha(z)\gamma(z)$. We give an example that corresponds to the following chosen coefficients:

$$a_{3n}=-\frac{1}{5},\ a_{3n+1}=\frac{1}{4},\ a_{3n+2}=0,\ b_{3n}=\frac{1}{2}\sqrt{\frac{2n+1}{3n+1}},\ b_{3n+1}=\frac{1}{2}\ \text{and}\ b_{3n+2}=\frac{1}{2}\sqrt{\frac{3n+1}{2n+1}}. \tag{19}$$

Thus, $\{A_1,A_2,A_3\}=\{-\frac{1}{5},\frac{1}{4},0\}$ and $\{B_1,B_2,B_3\}=\frac{1}{2}\left\{\sqrt{\frac{2}{3}},1,\sqrt{\frac{3}{2}}\right\}$. Figure 3a is a plot of the energy density obtained by substituting (18) into (9) then into (5) for large enough integer *N*. The three energy density bands are located within the following intervals:

$$E\in\left[-1.0360472248,-0.7207393595\right], \tag{20a}$$

$$E\in\left[-0.2739286325,+0.4365102472\right], \tag{20b}$$

$$E\in\left[+0.6495369776,+1.0446679919\right]. \tag{20c}$$

Figure 3b is a set of snap shot from the animation that shows the distribution of zeros (small filled circles) of the polynomials of the first kind that satisfy the recursion relation (2) with the recursion coefficients (19). Note the presence of three zeros within the two energy gaps. Only one of them, which is located in the middle of the left gap, is asymptotically stable against variations in the polynomial order and is common to all polynomials. This one corresponds to a bound state of the system with energy *E* = –0.499982389525. The polynomials $\left\{P_{3n+1}^{\mu}(E)\right\}$ have only one zero in the left gap that coincides with the bound states energy as $n\to\infty$. The polynomials $\left\{P_{3n+2}^{\mu}(E)\right\}$ have two zeros, one in each gap, that are stable asymptotically. The one in the left gap corresponds to the system's bound state. However, the one in the right gap corresponds to a bound state only for the $(3n+2)$–parity symmetric states (i.e., states that span the $\phi_{3n+2}(x)$ basis elements). At this energy, the asymptotic wavefunction is free from such states. On the other hand, the polynomials $\left\{P_{3n}^{\mu}(E)\right\}$ have two zeros both are located in the left gap and are stable asymptotically. Only one of them (the left one) coincides with the common bound state energy since it is present in all three sets of polynomials. The other one corresponds to a bound state for only the $(3n)$–parity symmetric states (i.e., states constructed using only $\phi_{3n}(x)$ basis elements). At this energy, the asymptotic wavefunction is free from such states.

# 5. The general construction

The most important ingredient in our construction of the energy density of states is the terminator $T(z)$, which is completely identified whence the three "*characteristic functions*" $\alpha(z)$, $\beta(z)$ and $\gamma(z)$ are given. Obviously, these functions depend only on the asymptotic coefficients $\left\{A_k, B_k\right\}_{k=1}^{K}$. The general construction starts by deciding on the number of density bands $K$ and their $2K$ boundaries. These boundaries uniquely determine the asymptotic limit $\left\{A_k, B_k\right\}_{k=1}^{K}$. The engineering part of the construction, which is the most laborious task, is to design several sets of recursion coefficients $\left\{a_n, b_n\right\}$ all with this same asymptotic limits and then choose the proper set that gives the desired shape of the individual density bands. This would entail proposing different expressions for $\left\{a_{Kn}, b_{Kn}\right\}$, $\left\{a_{Kn+1}, b_{Kn+1}\right\}$,..., $\left\{a_{Kn+K-1}, b_{Kn+K-1}\right\}$ all with the same asymptotic behavior $\lim_{n\to\infty}\left\{a_{Kn+k-1}, b_{Kn+k-1}\right\} = \left\{A_k, B_k\right\}$ for $k = 1,2,3,..,K$.

# 6. Discussion

In the three examples presented above, the bands of the energy density are all finite. However, our approach can still handle those with infinite bands. As a simple example, we could engineer a double band density as we did in section 3 but with $A_1 \pm A_2 = 2A_\pm$ being finite whereas $B_1 = B_2 = B$ such that $|B| >> |A_\pm|$. Consequently, the four boundaries of the two density bands are located at $E = A_+ \pm |A_-|$ and $E \approx \pm 2B$. Thus, as $B \to \infty$ we end up with a double band energy density centered at $E = A_+$ with a gap of width $2|A_-|$ and extending to $\pm\infty$. Specifically, we take the following recursion coefficients to replace (17)

$$a_{2n} = \gamma,\ a_{2n+1} = 1-\gamma,\ b_{2n} = \gamma\sqrt{2n+\alpha}, \text{ and } b_{2n+1} = \gamma\sqrt{2n+\beta}. \qquad (21)$$

The resulting energy density is shown in Figure 4 for a given set of physical parameters $\left\{\alpha,\beta,\gamma\right\}$. The left (right) side of the left (right) density band extends to minus (plus) infinity on the real energy axis, respectively.

An additional remark is about the type of energy polynomials used as expansion coefficients of the wavefunction. In all three examples above, those were taken as the polynomials of the first kind. Here, we give an example where the polynomials used are of the second kind. To be specific, we consider the case in section 2 for the single band energy density with the same recursion coefficients (14) but take $\hat{P}_0^\mu(E) = 1$ and $\hat{P}_1^\mu(E) = \left(-\frac{1}{2}E + 3a_0\right)/b_0$ as seeds for the recursion relation (2). Figure 5 shows the associated energy density of states $\hat{\rho}^\mu(E)$ obtained using Eq. (7). The difference between this plot and that given in Fig. 1a, which is associated with the polynomials of the first kind, is obvious.

## Figures Captions:

**Fig. 1a**: The single band energy density associated with system whose Hamiltonian matrix elements are given by Eq. (14) for $\alpha = 0.7$, $\beta = 0.5$ and $\gamma = -0.7$.

**Fig. 2a**: The double band energy density associated with system whose Hamiltonian matrix elements are given by Eq. (17) for $\alpha = 0.7$, $\beta = 0.8$ and $\gamma = 0.3$. The two energy bands are located in the energy intervals $E \in [-0.2762087348, 0.2938447187]$ and $E \in [0.7061552813, 1.2762087348]$.

**Fig. 3a**: The triple band energy density associated with system whose Hamiltonian matrix elements are given by Eq. (19). The three energy bands are located in the energy intervals (20).

**Fig. 1b**: Snap shots from the animation that shows the distribution of zeros (small filled circles) of the polynomials associated with the single band energy density of Fig. 1a. The vertical stem lines are at the boundaries of the band. The orders of the polynomials from top to bottom are 10, 20 and 30.

**Fig. 2b**: Snap shots from the animation that shows the distribution of zeros (small filled circles) of the polynomials associated with the double band energy density of Fig. 2a. The vertical stem lines are at the boundaries of the two bands. The left (right) column of the figure is for odd (even) polynomials with increasing degree from top to bottom. Note the presence of a single zero for odd polynomials located very close to the left edge of the energy gap.

**Fig. 3b**: Snap shots from the animation that shows the distribution of zeros (small filled circles) of the polynomials associated with the triple band energy density of Fig. 3a. The left, middle, and right columns of the figure are for the $P_{3n}(E)$, $P_{3n+1}(E)$, and $P_{3n+2}(E)$ polynomials, respectively. The degree of the polynomials increase from top to bottom. Note the presence of double, single, and double zeros within the two energy gaps for these respective polynomials. Only one of them, which is located in the middle of the left gap, is common to all of them and coincides with the energy of the system's bound state.

**Fig. 4**: The double band energy density associated with system whose Hamiltonian matrix elements are given by Eq. (21) for $\alpha = 1.0$, $\beta = 0.2$ and $\gamma = 0.8$. The energy gap is in the interval $E \in [0.2, 0.8]$ and the left (right) side of the left (right) density band extends to minus (plus) infinity.

**Fig. 5**: The single band energy density associated with system whose Hamiltonian matrix elements are given by Eq. (14) for $\alpha = 0.7$, $\beta = 0.5$ and $\gamma = -0.7$. This is identical to the system associated with Fig. 1a except that the energy polynomials used as expansion coefficients of the wavefunction are the second kind not the first kind with $\hat{P}_1^{\mu}(E) = \left(-\frac{1}{2}E + 3a_0\right)/b_0$.

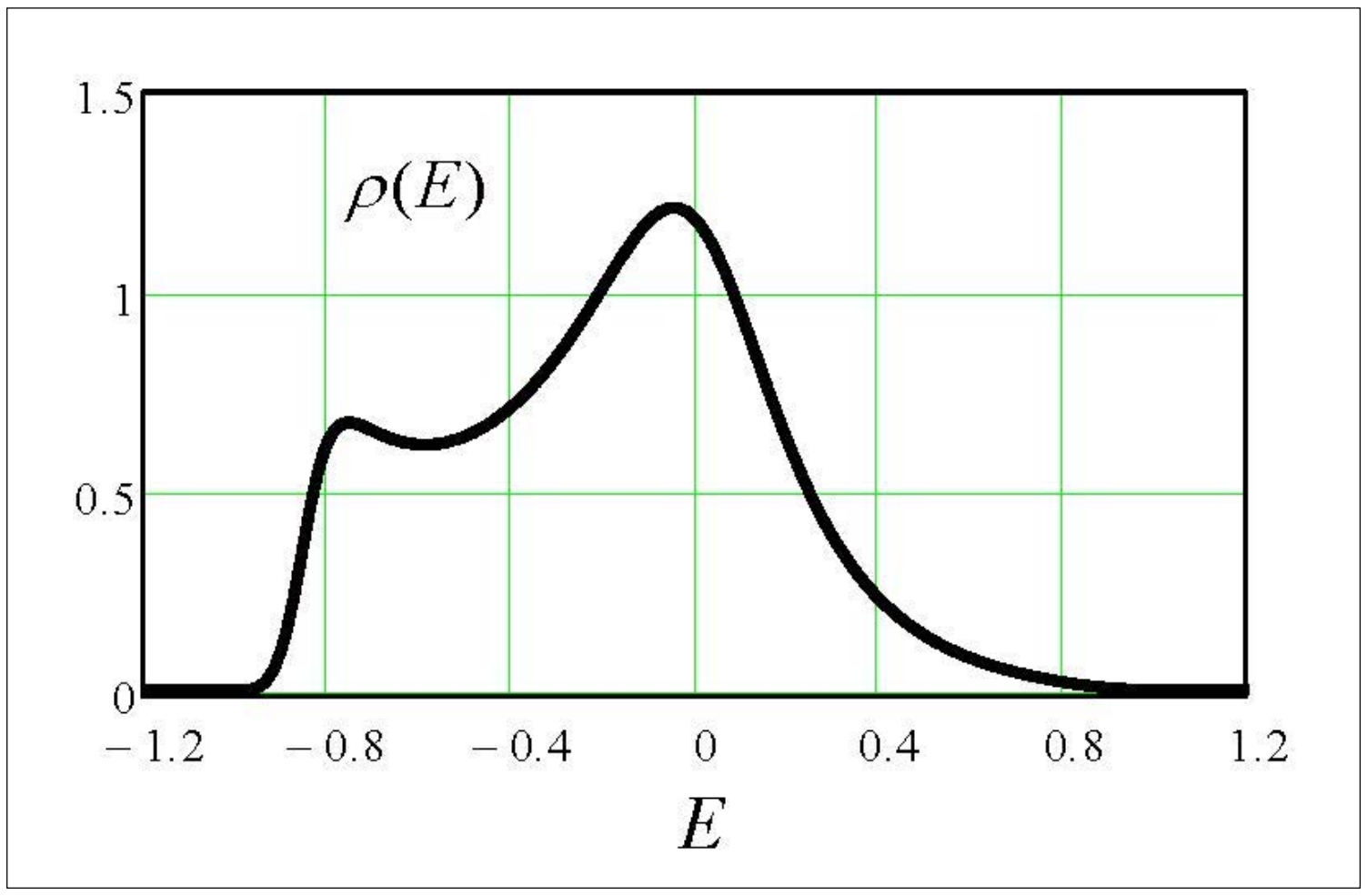

**Fig. 1a**

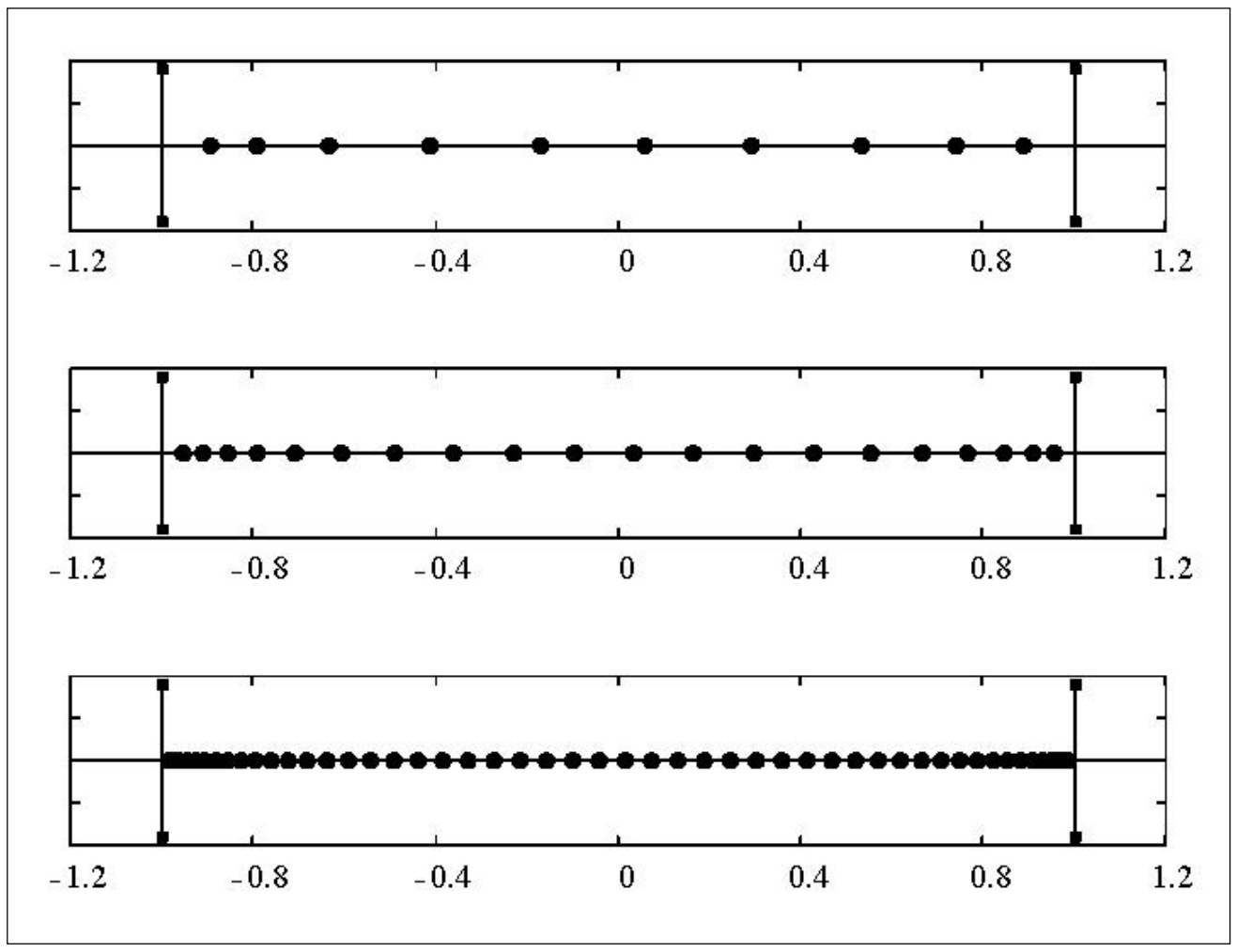

**Fig. 1b**

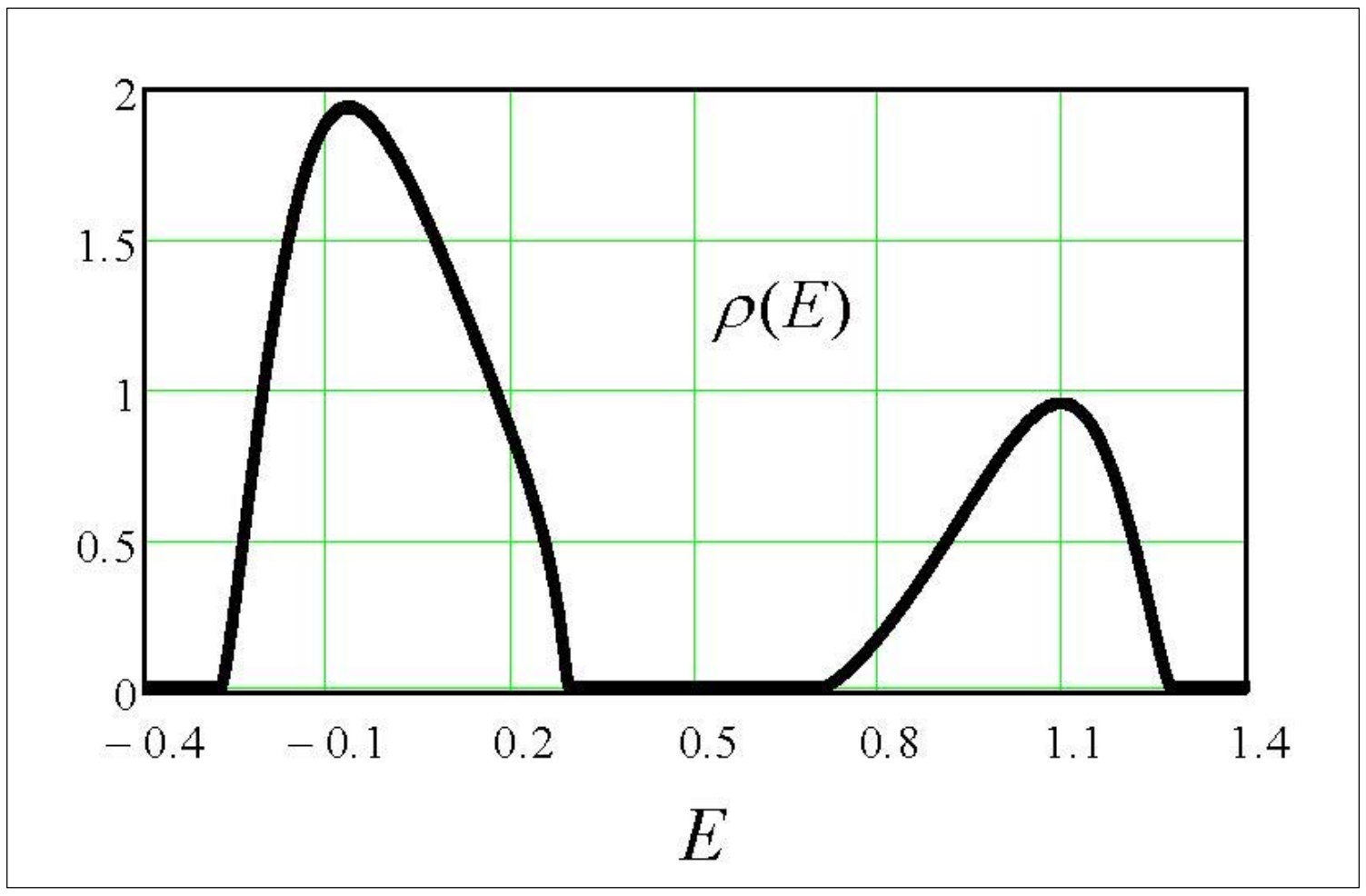

**Fig. 2a**

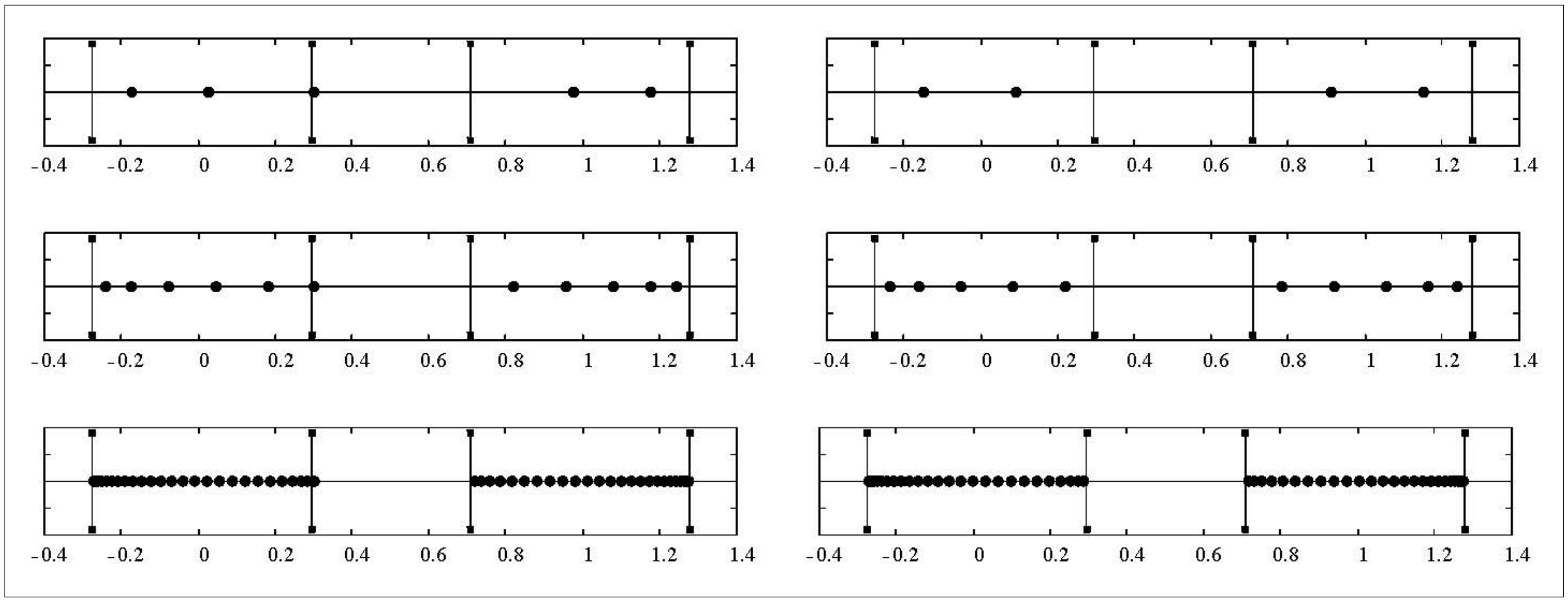

**Fig. 2b**

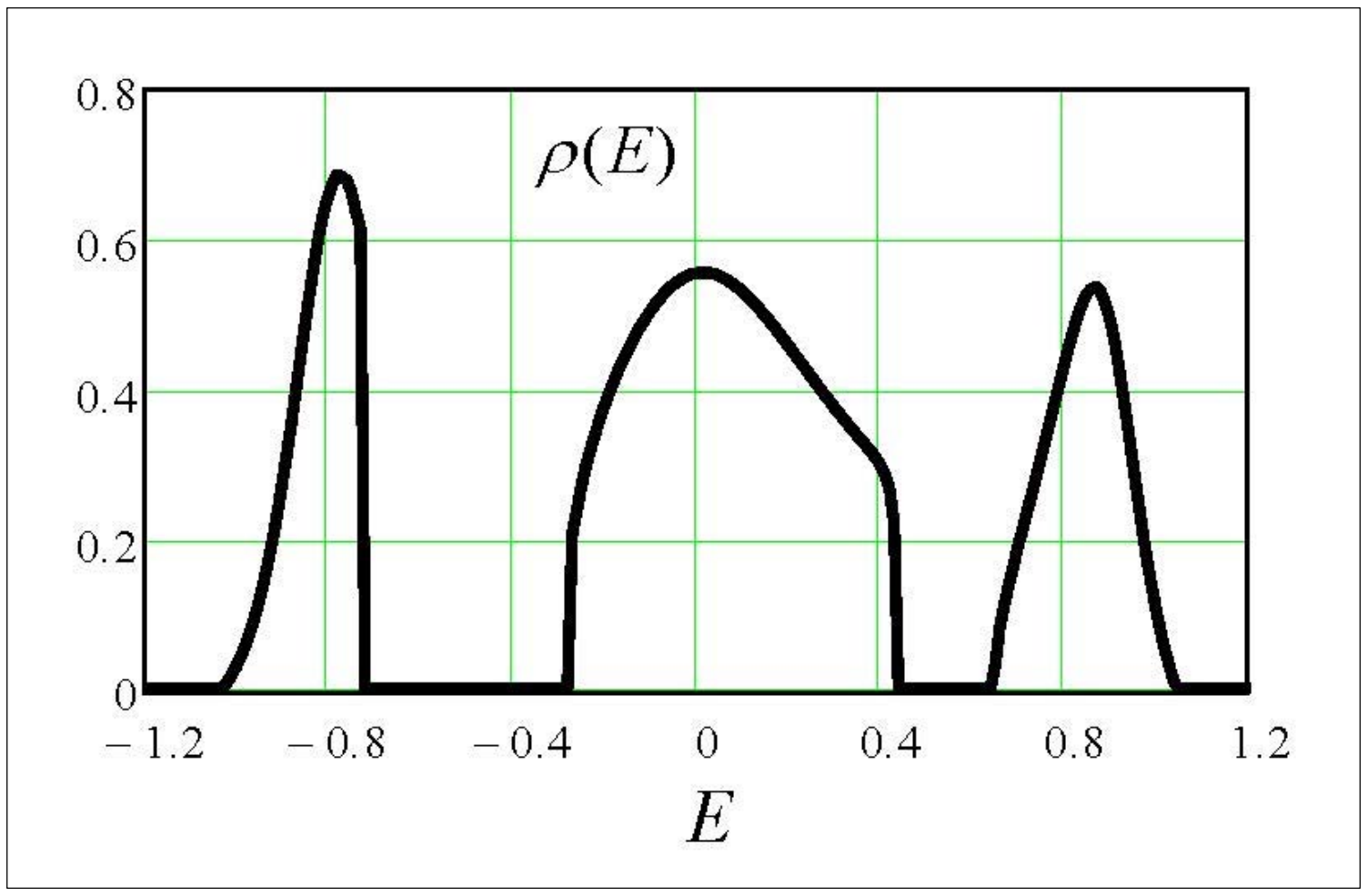

**Fig. 3a**

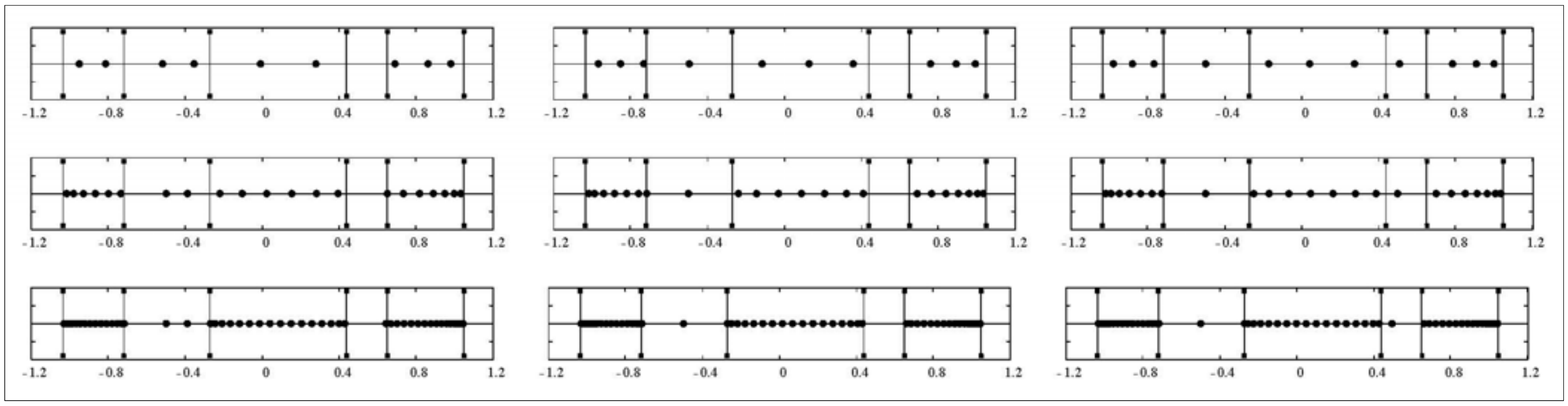

**Fig. 3b**

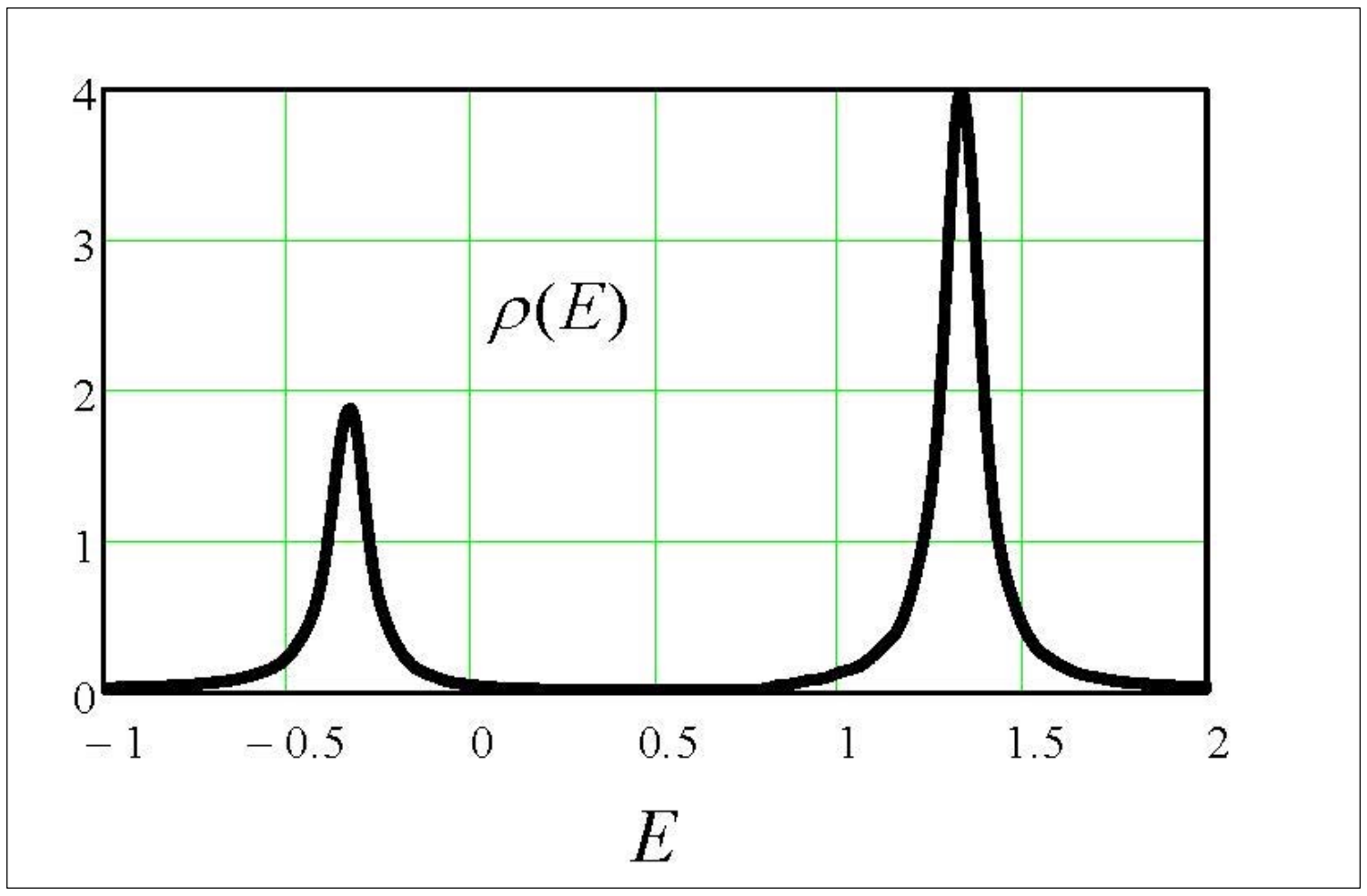

**Fig. 4**

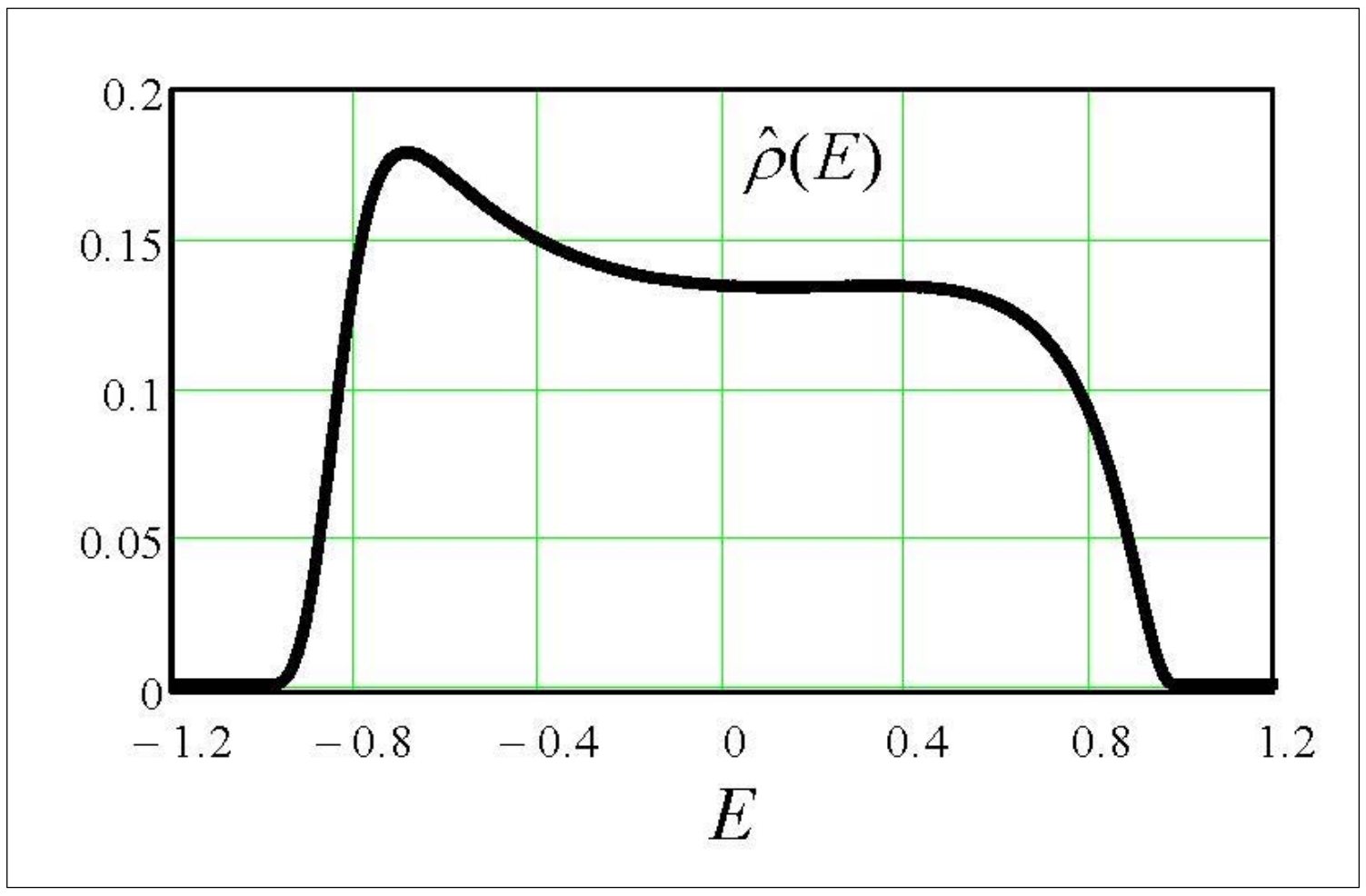

**Fig. 5**